\documentclass[]{spie}  

\usepackage{amsmath,amsfonts,amssymb,mathrsfs}
\usepackage{graphicx}
\usepackage[colorlinks=true, allcolors=blue]{hyperref}
\usepackage{astro_bib_macro} 

\title{Analytical model-based analysis of long-exposure images from ground-based telescopes}

\author{Lucie Leboulleux\supit{a}, Rapha\"{e}l Galicher\supit{a}, Eric Gendron\supit{a}, Pierre Baudoz\supit{a}, Gérard Rousset\supit{a}
\skiplinehalf
\supit{a} LESIA, Observatoire de Paris, Université PSL, CNRS, Sorbonne Université, Université de Paris, 5 place Jules Janssen, 92195 Meudon, France
}

\authorinfo{Further author information, send correspondence to Lucie Leboulleux: E-mail: lucie.leboulleux@obspm.fr, Telephone: 33 145077723}

\begin{document} 
\maketitle

\begin{abstract}

The search for Earth-like exoplanets requires high-contrast and high-angular resolution instruments, which designs can be very complex: they need an adaptive optics system to compensate for the effect of the atmospheric turbulence on image quality and a coronagraph to reduce the starlight and enable the companion imaging. During the instrument design phase and the error budget process, studies of performance as a function of optical errors are needed and require multiple end-to-end numerical simulations of wavefront errors through the optical system. 

In particular, the detailed analysis of long-exposure images enables to evaluate the image quality (photon noise level, impact of optical aberrations and of adaptive optics residuals, etc.). Nowadays simulating one long but finite exposure image means drawing several thousands of random frozen phase screens, simulating the image associated with each of them after propagation through the imaging instrument, and averaging all the images. Such a process is time consuming, demands a great deal of computer resources, and limits the number of parametric optimization.

We propose an alternative and innovative method to directly express the statistics of ground-based images for long but finite exposure times. It is based on an analytical model, which only requires the statistical properties of the atmospheric turbulence. Such a method can be applied to optimize the design of future instruments such as SPHERE+ (VLT)  or the planetary camera and spectrograph (PCS - ELT) or any ground-based instrument.


\end{abstract}

\keywords{Simulation, turbulence, error budget}

\section{INTRODUCTION}
\label{sec:INTRODUCTION}

The performance of an instrument is degraded by numerous factors, such as vibrations or thermal drifts of its different components, quasi-static or static aberrations, for instance due to a misalignment, and so on. In particular, the atmospheric turbulence\cite{Roddier81} impacts the quality of the images (Signal-to-Noise Ratio abbreviated SNR and angular resolution of the system) taken from ground-based telescopes. This phenomenon can be partly compensated by Adaptive Optics (AO) systems, but dynamic residuals remain in the final image and can vary at a high temporal frequency ($\simeq1\,$kHz).

The study of the impact of dynamic patterns on the resulting image has to be conducted during the instrument design and performance prediction. It requires numerical simulations of the instrument and of the images it provides for various configurations: design parameters, atmospheric turbulence conditions, stability, aberrations, misalignment, etc., the objective being to converge towards the most performant and stable design.

The image of the star through the instrument has to be simulated but not only in its core: faint objects (planets or circumstellar disks) are searched at a few angular resolution elements from the star, and accurate simulations of the star environment image are required. Several exposure times can be simulated: short, medium, and infinitely long exposure times.

In particular, infinitely long exposures completely average the dynamic speckle-boiling effects and provide a first infinitely smooth estimation of the image.  Analytical models have been proposed to simulate point spread function (PSF) with infinite exposure times: Roddier\cite{Roddier81, Roddier1999} sets up the formalism to compute the infinite exposure PSF from the phase power spectral density (PSD). It has later on been applied to AO-corrected images\cite{Jolissaint2002, Conan2005, Fetick2018, Fetick2019} and on post-coronagraph high-contrast images\cite{Sauvage2010, Herscovici-Schiller2016, Herscovici-Schiller2017}.

In this article, we define the "medium exposure" term for identifying an image that differs from the widely accepted terms of either short- or long-exposures one, as it stopped its convergence process somewhere in between those latter. The medium exposure image is a long, but finite exposure one and its simulation is crucial: it is more realistic since it corresponds to real on-sky observations and it enables a finer analysis of the image. Indeed ground-based instrument images tend to infinite exposure time images but also present additional random intensity variations of the star environment, ie. speckles that can hide a faint science object and are a main performance limitation. Estimating the amplitude of fluctuations of the PSF resulting from an image that has not been completely averaged by infinite exposures is then crucial to measure the high-contrast performance since it enables to test if the required performance, in particular in terms of SNR, can be reached to detect this faint companion, or to evaluate the minimum exposure time allowing its detection.


We can notice here that statistics of short or instantaneous PSFs can also be found in the litterature for AO-corrected images, without or with coronagraphy: Fusco \& Conan\cite{Fusco2004}, Aime \& Soummer\cite{Aime2004}, Soummer et al. \cite{Soummer2007, Soummer2007b}.

The classical method to simulate medium exposure images and their quality is based on the selection and propagation of multiple phase screens. In this proceeding, we propose an alternative method, based on the statistics of medium exposure images, to estimate their quality. In particular, we will need to analytically express the first two moments if such PSFs, meaning their mathematical expectation and variance. Unlike the classical method that requires a large number of random phase screens to simulate one medium exposure and extract its quality, our method directly computes its two first moments through their analytical expressions, which directly provides a knowledge of the SNR in the whole field with no need of generating multiple medium exposure images. 


This proceeding is organized as follows: in the section \ref{sec:ClassicalMethod} we present the classical method to simulate medium exposure time PSFs, in the section \ref{sec:Statistics} we analytically express the first two moments of medium exposure PSFs meaning their mathematical expectation and variance, and in the section \ref{sec:Conclusions} we report our conclusions and further studies.


\section{Classical simulation of medium exposure images}
\label{sec:ClassicalMethod}

The medium exposure time image is a long but finite exposure one, that stopped to converge between the short and the infinitely long exposure time images already studied in the literature. It is defined as $I_T(t) = \frac{1}{T}\int_{t-T}^{t} I(u) \, \mathrm{d}u$, where $T$ is the exposure time.

This section consists in reminders on the classical numerical simulation of $I_T(t)$ and is divided into three parts: \ref{sec:Instantaneous image formation}) the formation of an instantaneous image, \ref{sec:Principle}) the numerical simulation of a long exposure image, and \ref{sec:Remarks and conclusions}) some comments on the numerical burden of this method.

\subsection{Instantaneous image formation}
\label{sec:Instantaneous image formation}

We call $A$ the electric field in the entrance pupil plane of the system, $P$ the function that describes the entrance pupil, and $\phi(t)$ the instantaneous pupil phase at instant~$t$. The vector~$\vec{\xi}$ being the 2D-position vector in the pupil plane, we have the following equation:
\begin{equation} \label{eq:1}
    A(\vec{\xi}, t) = P(\vec{\xi})e^{i \phi(\vec{\xi}, t)} 
\end{equation}
For later developments in this proceeding, we assume that $\phi(\vec{\xi},t)$ is one realization of a random variable that follows a normal distribution with respect to time~$t$ and space~$\vec{\xi}$\cite{Roddier81}. We also consider that the mathematical expectation of these distributions is~$0$ and that the temporal standard deviation at position~$\vec{\xi}$ is $\sigma(\vec{\xi})$.

We consider that the science detector is in the focal plane of the telescope so that we can use the Fraunhofer approximation. The instantaneous amplitude~$A_f$ of the electric field in the detector plane is then given by:
\begin{equation} \label{eq:2}
    A_f(\vec{x},t) = \mathscr{F}\left[A\right](\vec{x},t)
\end{equation}
where $\vec{x}$ is the position vector in the detector plane and $\mathscr{F}\left[f\right]$ is the optical Fourier Transform of the function $f$. The instantaneous intensity~$I(\vec{x},t)$ in the image in visible or near-infrared light is proportional to the square modulus of the electric field so that:
\begin{equation} \label{eq:3}
    I(\vec{x},t) = \left \Vert A_f(\vec{x},t) \right \Vert ^2
\end{equation}
From a mathematical point of view, the instantaneous intensity~$I(\vec{x},t)$ is one realization of a random variable with spatial and temporal variations of its moments.

\subsection{Principle}
\label{sec:Principle}

In practice, the PSF is considered as instantaneous if its exposure time is shorter than the turbulence life time $t_0$, ie. the turbulent phase screen seen by the camera is frozen over $t_0$.

For an exposure time $T$ larger than $t_0$, the phase screen varies. We consider its variations as successive instantaneous screen realizations of the turbulent phase. For a turbulence lifetime $t_0$ short enough, these frozen screens can be seen as independent from each other. Under this hypothesis, simulating an image with an exposure time $T$ requires $N = \frac{T}{t_0}$ independent phase screens.

These frozen phase screens are one by one propagated through the optical system to generate instantaneous images with the process described in part \ref{sec:Instantaneous image formation} and the long exposure image corresponds to the integration over time of these time-discrete instantaneous images divided by the exposure time:
\begin{equation} \label{eq:4}
    \langle I(\vec{x}, t) \rangle_{T} = \frac{1}{N}\,\sum_{p=1}^{N} I(\vec{x},p\,t_0).
\end{equation}

\subsection{Remarks and conclusions}
\label{sec:Remarks and conclusions}

For ground-based instruments, the high temporal frequency of the turbulence (corrected or not by AO) imposes a high temporal resolution. The turbulence lifetime is typically of the order of a few milliseconds. If we set $t_0=2$ms, $N=30000$ instantaneous PSFs have to be computed to cover an exposure time $T=60$s.

In addition, current and coming ground-based imaging instruments are set up on very to extremely large telescopes, with diameters from 8 to 40m. These apertures have to be simulated at high spatial resolution. For telescopes such as the VLT (8m diameter primary mirror)\cite{Rousset2003, Beuzit2008} or the ELT (40m diameter)\cite{Kasper2008,Davies2010,Quanz2015}, this means very large arrays to compute and manipulate.

Between the number of realizations of the turbulent phase needed to simulate a long exposure image, the sizes of the arrays, and the number of atmospheric conditions that have to be tested\cite{Assemat2006}, the involved computational burden is enormous and completing a full tolerancing or design parameter exploration is extremely slow.

However, despite the demanding computational effort, such simulations are needed to design any instrument: GPI at the Gemini South Telescope \cite{Marois2008}, SPHERE at the Very Large Telescope \cite{Yaitskova2010, Carbillet2008}, MICADO \cite{Clenet2019, Clenet2018, Davies2018} at the Extremely Large Telescope \cite{Ferreira2018}.


\section{Statistical approach to long exposure images}
\label{sec:Statistics}

In this section, we use the work by Roddier \cite{Roddier81} to express the intensity of a medium exposure as a random variable and calculate the first two moments of this variable, ie. \ref{subsec:Mathematical expectation}) its mathematical expectation and \ref{subsec:Variance}) its variance. We also propose a few comments in the section~\ref{subsec:AlternativeMethod}.

\subsection{Mathematical expectation of the long exposure image}
\label{subsec:Mathematical expectation}

We call $\mathrm{E}[\langle I \rangle_{T}]$ the mathematical expectation of $\langle I \rangle_{T}$. We propose below a quick demonstration that $\mathrm{E}[\langle I \rangle_{T}] = \mathrm{E}[I]$, where $\mathrm{E}[I]$ corresponds to the infinite exposure time image. We also simplify the notation $I(p\,t_0)$ to $I(p)$:
\begin{equation} \label{eq:5}
\begin{split}
    \mathrm{E}[\langle I \rangle_{T}] & = \mathrm{E}[\frac{1}{N}\,\sum_{p=1}^{N} I(p)] \\
    &  = \frac{1}{N}\,\mathrm{E}[\sum_{p=1}^{N} I(p)] \\
    &  = \frac{1}{N}\,\sum_{p=1}^{N} \mathrm{E}[I]
\end{split}
\end{equation}
since $I(p)$ are independent variables. $\mathrm{E}[I]$ does not depend on $p$ so we directly obtain: 
\begin{equation} \label{eq:6}
    \mathrm{E}[\langle I \rangle_{T}] = \mathrm{E}[I]
\end{equation}
The expression of $\mathrm{E}[I] = \langle I \rangle_\infty$, the temporal mathematical expectation of $I(t)$, can be found in the literature\cite{Roddier81}:
\begin{equation} \label{eq:7}
    \langle  I \rangle_\infty (\vec{x}) = \mathrm{E}[I](\vec{x}) = \mathscr{F} \left[P \otimes P \times e^{-0.5 D_{\phi}}\right](\vec{x})
\end{equation}
where $P \otimes P$ is the pupil autocorrelation function and $D_{\phi}$ is the structure function of the phase $\phi$, defined as:
\begin{equation} \label{eq:8}
     D_{\phi}(\vec{\xi}) = \mathrm{E}\left[|\phi(\vec{\xi}'+\vec{\xi}) - \phi(\vec{\xi}') |^2 \right]
\end{equation}
The simulation of the infinite exposure time image~$\langle  I \rangle_\infty$ is thus quite straightforward, can easily be numerically calculated, and provides the first moment of the long and finite exposure time PSF.

\subsection{Variance of the long exposure image}
\label{subsec:Variance}

The objective of this section is to determine the temporal variance of~$\langle  I \rangle_T$, noted $\mathrm{Var}_t\left[\langle  I \rangle_T\right]$. From Eq.~\ref{eq:4}, we derive:
\begin{equation} \label{eq:9}
\begin{split}
    \mathrm{Var}_t\left[\langle  I \rangle_T (\vec{x})\right] & = \mathrm{Var}_t\left[\frac{1}{N}\,\sum_{p=1}^{N} I(\vec{x}, p)\right] \\
    & = \frac{1}{N^2}\,\mathrm{Var}_t\left[ \sum_{p=1}^{N} I(\vec{x}, p)\right]
\end{split}
\end{equation}
Since $(I(p))_{p \in [\![ 1,N ]\!] }$ is a set of independent variables, one can write:
\begin{equation} \label{eq:10}
\begin{split}
    \mathrm{Var}_t\left[\langle  I \rangle_T (\vec{x})\right] & = \frac{1}{N^2}\,\sum_{p=1}^{p=N} \mathrm{Var}_t\left[I\right](\vec{x}) \\
    & = \frac{1}{N}\,\mathrm{Var}_t\left[I\right](\vec{x})
\end{split}
\end{equation}
as the~$(I(p))_{p \in [\![ 1,N ]\!] }$ follow the same random variable. By definition of the variance:
\begin{equation} \label{eq:11}
\begin{split}
    \mathrm{Var}_t\left[I\right] & =\mathrm{E} \left[I^2\right]-\mathrm{E} \left[I\right]^2 \\
    & =\mathrm{E} \left[I^2\right]-I_\infty^2
\end{split}
\end{equation}
$E[I^2]$ can be expressed as follows:
\begin{equation} \label{eq:12}
    \mathrm{E}[I^2](\vec{x}) = \int \int \int \int P(\vec{\xi_1})P(\vec{\xi_2})P^*(\vec{\xi_3})P^*(\vec{\xi_4}) \mathrm{E}\left[e^{i (\phi(\vec{\xi_1}, t)+\phi(\vec{\xi_2}, t)-\phi(\vec{\xi_3}, t)-\phi(\vec{\xi_4}, t))}\right] e^{-i k \vec{x} . (\vec{\xi_1}+\vec{\xi_2}-\vec{\xi_3}-\vec{\xi_4})} d\vec{\xi_1} d\vec{\xi_2} d\vec{\xi_3} d\vec{\xi_4}
\end{equation}
In this expression, we focus on $\mathrm{E}\left[e^{i (\phi(\vec{\xi_1}, t)+\phi(\vec{\xi_2}, t)-\phi(\vec{\xi_3}, t)-\phi(\vec{\xi_4}, t))}\right]$. Since $\phi$ obeys a normal distribution,
\begin{equation} \label{eq:13}
   \mathrm{E}\left[e^{\displaystyle i \left(\phi(\vec{\xi_1}, t)+\phi(\vec{\xi_2}, t)-\phi(\vec{\xi_3}, t)-\phi(\vec{\xi_4}, t)\right)} \right] = e^{\displaystyle-\frac{1}{2}\, \alpha}
\end{equation}
where $\alpha = \mathrm{E}\left[ \left(\phi(\vec{\xi_1}, t)+\phi(\vec{\xi_2}, t)-\phi(\vec{\xi_3}, t)-\phi(\vec{\xi_4}, t)\right)^2 \right]$. Developing this expression provides:
\begin{equation} \label{eq:14}
\begin{split}
    \alpha & = \mathrm{E}\left[ \phi^2(\bold{\xi_1})+\phi^2(\bold{\xi_2})+\phi^2(\bold{\xi_3})+\phi^2(\bold{\xi_4}) \right]  \\
    & + \mathrm{E}\left[ 2 \phi(\bold{\xi_1})\phi(\bold{\xi_2}) + 2 \phi(\bold{\xi_3})\phi(\bold{\xi_4}) - 2 \phi(\bold{\xi_1})\phi(\bold{\xi_3}) - 2 \phi(\bold{\xi_1})\phi(\bold{\xi_4}) - 2 \phi(\bold{\xi_2})\phi(\bold{\xi_3}) - 2 \phi(\bold{\xi_2})\phi(\bold{\xi_4}) \right] \\
    & = 4 \mathrm{E}\left[ \phi^2(\bold{\xi_1}) \right] \\
    & + 2 \mathrm{E}\left[ \phi(\bold{\xi_1})\phi(\bold{\xi_2})\right] 
    + 2 \mathrm{E}\left[ \phi(\bold{\xi_3})\phi(\bold{\xi_4})\right] 
    - 2 \mathrm{E}\left[ \phi(\bold{\xi_1})\phi(\bold{\xi_3}) \right] 
    - 2 \mathrm{E}\left[ \phi(\bold{\xi_1})\phi(\bold{\xi_4}) \right] 
    - 2 \mathrm{E}\left[ \phi(\bold{\xi_2})\phi(\bold{\xi_3}) \right] 
    - 2 \mathrm{E}\left[ \phi(\bold{\xi_2})\phi(\bold{\xi_4}) \right] 
\end{split}
\end{equation}
because of a spatial translational invariance. Then, we use $D_{\phi}(\vec{\xi}) = 2\, \mathrm{E}\left[ \phi^2(\vec{\xi},t)  \right]  - 2\, \mathrm{E}\left[\phi(\vec{\xi},t) \phi(\vec{\xi}',t) \right] $ to obtain:
\begin{equation} \label{eq:15}
    \alpha = D_\phi\left(\vec{\xi_3}-\vec{\xi_1}\right) + D_\phi\left(\vec{\xi_4}-\vec{\xi_1}\right) + D_\phi\left(\vec{\xi_3}-\vec{\xi_2}\right) + D_\phi\left(\vec{\xi_4}-\vec{\xi_2}\right) - D_\phi\left(\vec{\xi_2}-\vec{\xi_1}\right) - D_\phi\left(\vec{\xi_4}-\vec{\xi_3}\right)
\end{equation}
where the phase structure function has been defined in Eq.~\ref{eq:8}. Hence, combining Eq.~\ref{eq:10}, \ref{eq:11}, \ref{eq:7}, \ref{eq:12}, \ref{eq:13}, and \ref{eq:15} provides the variance:
\begin{equation} \label{eq:16}
\begin{split}
    & \mathrm{Var}_t\left[\langle  I \rangle_T \right] = \frac{\mathrm{E} \left[I^2\right]-\mathrm{E}[I]^2}{N}, \\
    & \textrm{with } \mathrm{E}[I] = \mathscr{F} \left[P \otimes P \times e^{-0.5 D_{\phi}}\right], \\
    & \textrm{\hspace{0.8cm}} \mathrm{E} \left[I^2\right] (\vec{x}) = \int \int \int \int P(\vec{\xi_1})P(\vec{\xi_2})P^*(\vec{\xi_3})P^*(\vec{\xi_4}) e^{\displaystyle-\frac{1}{2}\, \alpha} e^{-i k \vec{x} . (\vec{\xi_1}+\vec{\xi_2}-\vec{\xi_3}-\vec{\xi_4})} d\vec{\xi_1} d\vec{\xi_2} d\vec{\xi_3} d\vec{\xi_4}, \\
    & \textrm{and } \alpha = D_\phi\left(\vec{\xi_3}-\vec{\xi_1}\right) + D_\phi\left(\vec{\xi_4}-\vec{\xi_1}\right) + D_\phi\left(\vec{\xi_3}-\vec{\xi_2}\right) + D_\phi\left(\vec{\xi_4}-\vec{\xi_2}\right) - D_\phi\left(\vec{\xi_2}-\vec{\xi_1}\right) - D_\phi\left(\vec{\xi_4}-\vec{\xi_3}\right)
\end{split}
\end{equation}

\subsection{Summary and remarks}
\label{subsec:AlternativeMethod}

The long exposure image~$\langle I \rangle_T$ can be seen as one realization of a random variable. Its first two moments have been analytically expressed in the previous section:

- its mathematical expectation~$E[\langle I \rangle_T]$ also corresponds to $\langle  I \rangle_\infty$ given by Eq.~\ref{eq:7},

- its variance~$\mathrm{Var}[\langle I \rangle_T]$ can be calculated with the model of Eq.~\ref{eq:16}.


$\langle I \rangle_\infty$ only provides an approximation of the image~$\langle I \rangle_T$ recorded with the finite exposure time $T$. The temporal variation of~$\langle I \rangle_T$ appears in its variance~$\mathrm{Var}[\langle I \rangle_T]$. We notice that this variance decreases as $t_0/T$ (first line of Eq.~\ref{eq:16}), meaning that the long exposure image converges as expected towards $\langle  I \rangle_\infty$ when the exposure time increases. 

The main advantage of this analytical expression of the variance is that it provides a direct knowledge of the SNR in the whole field with no need of generating multiple medium exposure images. 

\section{Conclusions and perspectives}
\label{sec:Conclusions}

In this proceeding, we focused on medium exposure time images, with the objective of accurately estimating the performance of future ground-based imaging instruments. The estimation of the quality (SNR) of medium exposure images is necessary since they correspond to realistic on-sky observations. They also enable the study of dynamic patterns not fully averaged, unlike infinite exposure image models, that can prevent the detection of a faint object in images that show stellar speckles. 

Usually, a medium exposure time image is simulated by averaging a large number of instantaneous images and its quality is estimated afterwards. We propose another approach deriving the mathematical distribution of the long exposure PSF. This method is based on the analytical expressions of the first two moments of such PSFs, meaning its mathematical expectation and its variance over time.

Analytical models of the infinite exposure time image already exist and provide the first moment of the long exposure image. In this proceeding, we derived a model for its second moment, ie. its temporal variance. Both expressions require only the statistics of the wavefront, expressed through the phase structure function. Knowing its mathematical expectation and its variance, the quality of the image and in particular its SNR in the whole detector can be generated while avoiding multiple medium exposure images made of hundreds to thousands of short exposure images.

To go further, we could propose an alternative method to simulate medium exposure images, derived from their analytical statistics only. In addition to the mathematical expectation and the variance, this would require to also compute the covariance matrix of the long exposure image and demand more analytical developments.

This analytical and statistical approach can be used on various ground-based imaging instruments, equipped or not with adaptive optics systems, such as SPHERE+ at the VLT, PCS or MICADO at the ELT. Further developments can provide an extension of the analytical models presented in this paper to coronagraphic instruments dedicated to high-contrast imaging.

\acknowledgments 
This work is supported by the IRIS Origines et Conditions d’Apparition de la Vie (OCAV) project of PSL Idex under the program Investissements d’Avenir with the reference ANR-10-IDEX-0001-02 PSL.

\bibliography{bib}
\bibliographystyle{lyot_spiebib}

\end{document}